\documentclass[12pt,paper]{JHEP3} 
\usepackage{graphicx}
\usepackage{amssymb}

\newcommand{\be}{\begin{equation}}
\newcommand{\ee}{\end{equation}}
\newcommand{\ba}{\begin{eqnarray}}
\newcommand{\ea}{\end{eqnarray}}

\newcommand{\im}{\mbox{Im}}

\preprint{LU TP 04-19\\
hep-ph/0404150\\
April 2004}

\title{$\pi K$ Scattering in Three Flavour ChPT\thanks{Supported
in part by the European Union TMR
network, Contract No. HPRN-CT-2002-00311  (EURIDICE)}
}
\vfill
\author{Johan Bijnens, Pierre Dhonte\\
Department of Theoretical Physics 2, Lund University,\\
S\"olvegatan 14A, S 223-62 Lund, Sweden}
\author{Pere Talavera\\
Departament de F\'\i sica i Enginyeria Nuclear, Universitat Polit\`ecnica
de Catalunya,\\Jordi Girona 1-3, E 08034 Barcelona, Spain}

\abstract{
We present the scattering lengths for the $\pi K$ processes in
the three flavour Chiral Perturbation Theory (ChPT) framework at 
next-to-next-to-leading order (NNLO). The calculation has been performed
analytically but we only include analytical results for the
dependence on the low-energy constants (LECs) at NNLO due to the size
of the expressions. These results, together with resonance estimates
of the NNLO LECs are used to obtain constraints on the Zweig rule
suppressed LECs at NLO, $L_4^r$ and $L_6^r$. Contrary to expectations
from NLO order calculations we find them to be compatible with zero. 
We do a preliminary study of combining the results from $\pi\pi$ scattering,
$\pi K$ scattering and the scalar form-factors and find only a marginal
compatibility with all experimental/dispersive input data.}


\begin{document}

\section{Introduction}

Effective Lagrangians have become a widely used tool in understanding
physics involving a mass gap in the spectrum. They can be used
in theories in a weakly coupled regime but with unknown underlying physics
(as is the case in the Higgs sector of the standard model)
 or in theories with a strongly coupled regime where the usual
perturbation formalism breaks down. Our interest will be
focused on Quantum Chromo Dynamics (QCD). It is
a well established theory at high energy where our theoretical knowledge and
the experimental outcome agree with rather good accuracy. 
At low energy the situation is less
satisfactory because the theory becomes strongly coupled and non-perturbative,
standard perturbative methods can not be applied.
One of the immediate differences is given by the degrees of freedom
at low and high energy.
At  the former they are characterised by hadrons while in the latter
they are better described in terms of the fundamental interacting quarks and
gluons. 

A suitable method to tackle phenomenology at low energy in the mesonic sector,
besides direct numerical  computation as done in lattice QCD,
is to use the fact that QCD possesses an almost exact symmetry.
One can then rely on these symmetries 
and their breaking pattern using an effective Lagrangian method.
We will use the chiral symmetry present in the QCD Lagrangian in the limit
of massless quarks. The use of this symmetry and the effective Lagrangian method
is now known as Chiral Perturbation Theory (ChPT).
It was introduced by Weinberg \cite{Weinberg} and systematized
by Gasser and Leutwyler \cite{GL1} for the case of the light up and down quarks
as well as for the case where the strange quark is treated as light
in addition \cite{GL2}.
They performed a basic set of next-to-leading order (NLO) calculations
allowing a first determination of all low-energy constants at NLO, the
$L_i^r$, invoking the Zweig rule to set $L_4^r=L_6^r\approx0$.
It was hoped this could be tested in $K_{\ell4}$ easily, but it
turned out, when the explicit calculations were performed, that these
only had very suppressed contributions to the
form-factors \cite{Bijnenskl4,Riggenbach}.

One question is how to order the various terms in the chiral
expansions. The proposal
used by most people is to count energies and momenta as a small parameter
of order $p$ and the quark masses as order $p^2$. Alternative countings,
taking the quark masses also as order $p$ are possible,
see \cite{Stern1} and references therein. Combining
the two-flavour two-loop calculations of $\pi\pi$ scattering
\cite{BCEGS1,BCEGS2} and the pion
scalar form-factor \cite{BCT} with the Roy equation analysis \cite{ACGL}
it could be shown that in the two flavour case the correct counting was
the standard one \cite{CGL1,CGL2} using the recent determination of the
pion scattering length from $K_{\ell4}$ \cite{Pislak1,Pislak2}.

The up and down quark masses are much smaller than the strange quark mass.
The question thus remains whether three flavour ChPT itself converges
sufficiently well to be of practical use and whether alternative countings
of contributions involving the strange quark mass need to be used.
This possibility is discussed in the recent work
\cite{Stern2,Stern3}. The argument is that disconnected
loop contributions from strange quarks, via kaons and etas, can be large,
making a convergent three flavour ChPT difficult to achieve in the usual
sense \cite{Stern3,LJP}. 
Answering this question is part of the larger problem of whether the strange
quark can be considered a light quark.
This was part of the motivation behind the recent work
in three flavour ChPT at NNLO on $\pi\pi$ scattering
 \cite{BDT} and the various scalar form-factors \cite{BD}.
Earlier calculations of the pseudoscalar
masses and decay constants, see Ref.~\cite{ABT1}
and references therein, showed the possibility of this behaviour.
The various vector form-factors calculated did not seem to have problems
with convergence \cite{Post1,Post2,BT1,BT2}.

Work on $\pi K$ scattering began soon after Weinberg's calculation of
$\pi\pi$ scattering \cite{Weinbergpipi}. The earliest reference known to us
is \cite{Griffith}. During the 1970s there was a dedicated series of
experiments culminating in the review by Lang \cite{Langreview}.
These were used extensively together with dispersion relations
and crossing symmetry in \cite{Lang} and \cite{JN}.

The development of ChPT led after some time to the calculation of the
$\pi K$ scattering amplitude to NLO \cite{BKM,BKM2}. There were also
earlier attempts at unitarization of current algebra for
this process. An example can be found in the discussion in 
\cite{SaBorges1,SaBorges2} and references therein.
Other recent related works are the various attempts at putting in
resonances in this process starting with \cite{BKM3}. Approaches involving
resummations can be found in \cite{JOP} and \cite{MO}. An alternative
approach is to consider the kaon as heavy and treat only the pion as a
Goldstone boson. This is known as heavy kaon ChPT. 
The applications to $\pi K$ scattering
can be found in \cite{Roessl} and \cite{FKM}. Unfortunately this approach has
many free parameters and does not allow to connect $\pi K$ scattering to other
processes. It does, however, have the possibility of being applicable
even if standard ChPT does not work.

On the dispersive side, the analyses of \cite{Lang} were slowly updated to
get at a determination of the $p^4$ LECs. The first work was \cite{AB}
and the full analysis has recently become available \cite{BDM}.
In the mean time, the isospin breaking corrections to $\pi K$ scattering
at NLO have been evaluated in \cite{Nehme1,Nehme2,KM1,KM2}. Since we
work in the isospin limit and the dispersive calculation of \cite{BDM} has been
performed in the same approximation we do not discuss these works further.

In this paper we calculate the standard ChPT expression for $\pi K$
scattering to next-to-next-to-leading-order. 
A large number of calculations to this order exist
in three flavour ChPT and we have thus been able to determine most
$p^4$ LECs with this precision after making some assumptions on the values
of the $p^6$ constants. In earlier work it has been found that the Zweig rule
suppressed constants $L_4^r$ and $L_6^r$ could be sufficiently different
from zero that the scenario of large corrections due to disconnected
strange quark loops was not ruled out. Pushing this calculation to NNLO
allows then to perform this comparison at the same footing as all the
other $p^4$ LECs. Earlier attempts at using $K_{\ell4}$ had led to rather
large errors for these constants, e.g.  
$L_4^r = (-0.2\pm0.9)\,10^{-3}$\cite{ABT4}.
The work on $\pi\pi$ scattering \cite{BDT} and scalar form-factors \cite{BD}
at NNLO order gave an indication that the region with
$L_4^r\approx 0.45\,10^{-3}$ was preferred. This fitted with the NLO $\pi K$
work done earlier \cite{BDM}. As discussed below, contrary to our expectations,
the results from $\pi K$ scattering at NNLO are more indicative of a smaller
value for $L_4^r$.

This paper is organized as follows. In Sect.~\ref{chpt} we give a very short
overview of ChPT and the references for NNLO techniques.
In Sect.~\ref{piKgeneral} we discuss a few  general properties of the
$\pi K$ scattering
amplitude.
Sect.~\ref{analytical} gives an overview of our main result, the
calculation of the $\pi K$ scattering amplitude to NNLO in three flavour ChPT.
We also present here some plots showing the importance of the various
contributions. The inputs we use to do the numerical analysis are
described in Sect.~\ref{inputs}. The main numerical analysis is
presented in Sect.~\ref{numerical} and we give our conclusions
in Sect.~\ref{conclusions}.

\section{Chiral Perturbation Theory}
\label{chpt}

ChPT is the effective field theory for QCD at low energies introduced in
 \cite{Weinberg,GL1,GL2}. Introductory references are
in Ref.~\cite{chptlectures}. 
The usual expansion is
in quark masses and meson momenta generically labeled $p$ and
assumes $m_q\sim p^2$.
The Lagrangian for the strong and semi leptonic mesonic sector
to NNLO can be written as
\be
{\cal L} = {\cal L}_2+ {\cal L}_4+{\cal L}_6\,,
\ee
where the subscript refers to the chiral order.
The lowest order Lagrangian is
\be
\label{L2}
{\cal L}_{2} = \frac{F_0^2}{4} \langle u_\mu u^\mu + \chi_+ \rangle \,.
\ee
The mesonic fields enter in a non-linear fashion via
$u = \exp\left( i M /(F_0 \sqrt{2})\right)$, with $M$ parametrising
the pseudoscalar fields.
The quantity $u_\mu$ also contains the external vector ($v_\mu$)
and axial-vector ($a_\mu$)
currents
\be
\label{covariant}
u_\mu = i 
(u^\dagger \partial_\mu u - \partial_\mu u u^\dagger 
 -i u^\dagger r_\mu u + i u l_\mu u^\dagger)\,,
\quad
l_\mu(r_\mu) = v_\mu -(+) a_\mu\,.
\ee
The scalar ($s$) and pseudo scalar ($p$) currents are contained in
\be
\chi_\pm = u^\dagger \chi u^\dagger \pm u \chi^\dagger u\,,\quad
\chi = 2 B_0\left(s+ip\right)\,.
\ee
The $p^4$ or NLO Lagrangian, ${\cal L}_4$, was introduced in Ref.~\cite{GL2}
and is of the general form
\be
\label{L4}
{\cal L}_{4} = \sum_{i=1}^{12} L_i P_i
= \ldots
+L_4 \langle u^\mu  u_\mu  \rangle \langle \chi_+ \rangle 
+L_6 \langle \chi_+ \rangle^2 \nonumber
+\ldots
\,.
\ee
We have explicitly shown two terms with chiral symmetry breaking 
which in addition present a double flavour trace structure, which indicates
that these two terms are suppressed by the Zweig rule.

The schematic form of the NNLO Lagrangian in the three flavour case is
\be
\label{L6}
{\cal L}_6 = \sum_{i=1,94} C_i\,O_i
\ee
and we  refer to \cite{BCE} for their explicit expressions.

The ultra-violet divergences produced by loop diagrams of order
$p^4$ and $p^6$ cancel in the process of renormalization with the divergences
extracted from the low-energy constants $L_i$'s and $C_i$'s.
We use dimensional regularization and
the modified minimal subtraction
$(\overline{MS})$ version usually used in ChPT.
An extensive description of the
regularization and renormalization procedure
can be found in Refs.~\cite{BCEGS2} and \cite{BCE2}.
The divergences are known in general \cite{BCE2,GL1,GL2,BCE3} and their cancellation
is a check on our calculation. 

\section{The $\pi K$ amplitude: general properties}
\label{piKgeneral}

The $\pi K$ scattering amplitude in isospin channel $I$
can be written as
\be
A(\left(\pi(p_1)K(p_2)\to\pi(p_3)K(p_4)\right)
= T^I(s,t,u)\,.
\ee
$s,t,u$ are the usual Mandelstam variables
\be
s = (p_1+p_2)^2\,,\quad t=(p_1-p_3)^2\quad\mbox{and}\quad u=(p_1-p_4)^2\,.
\ee
There are two possible isospin combinations $I=1/2$ and $I=3/2$. These
two are related via $s\leftrightarrow u$ crossing which yields
\be
T^{\frac{1}{2}}(s,t,u) =
 -\frac{1}{2}T^{\frac{3}{2}}(s,t,u)+\frac{3}{2}T^{\frac{3}{2}}(u,t,s)\,.
\ee
We also define the crossing symmetric amplitudes $T^+$ and $T^-$ as,
\be
T^{\pm}(s,t,u) =
 \frac{1}{2}\left(\pm T^{\frac{3}{2}}(s,t,u)+ T^{\frac{3}{2}}(u,t,s)\right)\,.
\ee
These amplitudes can be calculated most easily from the purely $I=3/2$
process $\pi^+ K^+\to\pi^+ K^+$.

In order to describe
scattering kinematics it is convenient
to introduce the variable
\be
q_{\pi K}^2 = \frac{s}{4}\left(1-\frac{(m_K+m_\pi)^2}{s}\right)
 \left(1-\frac{(m_K-m_\pi)^2}{s}\right)\,.
\ee
The kinematical variables $t,u$ can be expressed in terms of $s$ 
and $\cos\theta$ as
\be
t = -2 q_{\pi K}^2(1-\cos\theta)\,,\quad
u = -s-t+2 m_K^2+2 m_\pi^2\,.
\ee

The various amplitudes are expanded in partial waves via
\be
T^I(s,t,u) = 
16\pi\sum_{\ell=0}^\infty (2\ell+1) P_\ell(\cos\theta) t^I_\ell(s)\,.
\ee
Near threshold these can be expanded in a Taylor series 
\be
\label{defaij}
t^I_\ell(s) \equiv \frac{1}{2}\sqrt{s}
q_{\pi K}^{2\ell}\left(a^I_\ell + b^I_\ell q_{\pi K}^2 
+ {\cal O}(q_{\pi K}^4)\right)\,,
\ee
defining the threshold parameters $a^I_\ell$ and $b^I_\ell$.

Below the inelastic threshold the partial waves satisfy
\be
\im t^I_\ell(s) = \frac{2 q_{\pi K}}{\sqrt{s}} 
\left|t^I_\ell(s)\right|^2\,.
\ee
In ChPT the inelasticity only starts at order $p^8$.
In this regime the partial waves can be written in terms of the phase-shifts
as
\be
t^I_\ell(s) = \frac{\sqrt{s}}{2 q_{\pi K}}\,\frac{1}{2i}\left\{
e^{2i\delta^I_\ell(s)}-1\right\}\,.
\ee

The amplitudes are often expanded around the point $t=0$, $s=u$
via
\ba
\label{defci}
T^+(s,t,u) &=& \sum_{i,j=0}^\infty c^+_{ij} t^j \nu^{2j}\,, 
\nonumber\\
T^-(s,t,u) &=& \sum_{i,j=0}^\infty c^-_{ij} t^j \nu^{2j+1}\,, 
\ea
where $\nu = (s-u)/(4 m_K)$ and the $c^\pm_{ij}$ are referred to
as subthreshold expansion parameters. They are normally quoted in units of the
relevant power of $m_{\pi^+}$. We will always list these parameters in
increasing order corresponding to powers of $t$ and $s-u$.

The $p^2$ result for the amplitude 
has only a few nonzero items. As a consequence
the imaginary parts for all other partial waves starts only at order $p^8$.
This allows the amplitude to be written in the form
\be
\label{defGi}
T^{\frac{3}{2}}(s,t,u) = G_1(s)+G_2(t)+G_3(u)+(s-u) G_4(t)+(s-t) G_5(u)\,.
\ee 
The functions $G_i(s)$ have a polynomial ambiguity due to
$s+t+u=2 m_\pi^2+ 2 m_K^2$. 
The functions $G_i(s)$ have various singularities. $G_1$,  $G_3$ and $G_5$
contain singularities from the $\pi K$ and $\eta K$ intermediate states
and $G_2$ and $G_4$ from the possible nonstrange two meson intermediates.
The precise relation with the various singularities can be found in \cite{AB}.

\section{ChPT results}
\label{analytical}

\subsection{Results at order $p^2$}

The lowest order result is very simple and corresponds to
\be
c^+_{00} = 0,\quad c^+_{10} = \frac{1}{4 F_\pi^2}
\quad\mbox{and}\quad c^-_{00} = \frac{m_K}{F_\pi^2}\,.
\ee
All higher terms vanish. This was initially performed 
using current algebra methods
\cite{Griffith}.

\subsection{Results at order $p^4$}

The next-to-leading
calculation was first performed in \cite{BKM,BKM2}. 
We present it here in a slightly different form,
but the final expression given in that reference agrees with ours up to
terms of order $p^6$. Note that as mentioned also in \cite{AB} there are some
misprints in the formulas in \cite{BKM}.

We present the analytical expressions for the $G_i$ functions
defined in (\ref{defGi}) and where the polynomial part was isolated in a 
function $G_6(s,t,u)$. We also use $\Delta = m_K^2-m_\pi^2$ and 
$\Sigma=m_K^2+m_\pi^2$.

\begin{eqnarray}
F_\pi^4\, G_1(s) &=&\overline{B}_{}(m_\pi^2,m_K^2,s)\,(-1/2\,s\,\Sigma+1/4\,s^2+1/4\Sigma^2)\,,
\\
F_\pi^4\, G_2(t) &=&\overline{B}_{}(m_\pi^2,m_\pi^2,t)\,(1/16\,t\,\Delta-1/16\,t\,\Sigma+1/4\,t^2)\nonumber\\&&
+\overline{B}_{}(m_K^2,m_K^2,t)\,(3/16\,t^2)\nonumber\\&&
+ \overline{B}_{}(m_\eta^2,m_\eta^2,t)\,( 1/24\,m_\eta^2\,\Delta - 1/24\,m_\eta^2\,\Sigma - 1/16\,t\,\Delta
\nonumber\\ &&\hspace{1cm}+ 1/16\,t\,\Sigma+ 1/72\,\Delta\,\Sigma - 1/144\,\Delta^2 - 1/144\,\Sigma^2 )\,,
\\
F_\pi^4 G_3(u) &=&\overline{B}_{}(m_\pi^2,m_K^2,u)\,(-3/8\,u\,\Sigma+11/32\,u^2+1/8\Delta\Sigma
\nonumber\\&&\hspace{3cm}-7/32\Delta^2+1/8\Sigma^2)\nonumber\\&&
+\overline{B}_{}(m_K^2,m_\eta^2,u)\,(  - 1/16\,m_\eta^2\,u - 41/288\,m_\eta^2\,
\Delta + 1/32\,m_\eta^2\,\Sigma 
\nonumber\\&&\hspace{3cm}+ 1/96\,m_\eta^4 + 5/96\,u\,\Delta 
- 3/32\,u\,\Sigma + 3/32\,u^2
\nonumber\\&&\hspace{3cm}+ 31/192\,\Delta\,\Sigma + 493/3456\,\Delta^2 
+ 3/128\,\Sigma^2 )
\nonumber\\&&
+\overline{B}_{1}(m_\pi^2,m_K^2,u)\,(-1/4\Delta\Sigma-3/8\Delta^2)\nonumber\\&&
+\overline{B}_{1}(m_K^2,m_\eta^2,u)\,(  - 1/8\,m_\eta^2\,\Delta 
- 3/16\,\Delta\,\Sigma - 13/48\,\Delta^2 )\nonumber\\&&
+\overline{B}_{21}(m_\pi^2,m_K^2,u)\,(3/8\Delta^2)\nonumber\\&&
+\overline{B}_{21}(m_K^2,m_\eta^2,u)\,(3/8\Delta^2)\,,
\\
F_\pi^4\, G_4(t) &=&\overline{B}_{}(m_\pi^2,m_\pi^2,t)\,(-1/24\,t\,-1/12\Delta+1/12\Sigma)\nonumber\\&&
+\overline{B}_{}(m_K^2,m_K^2,t)\,(-1/48\,t\,+1/24\Delta+1/24\Sigma)\,,
\\
F_\pi^4\, G_5(u) &=&\overline{B}_{}(m_\pi^2,m_K^2,u)\,(-1/32u-1/32\Delta+1/16\Sigma)\nonumber\\&&
+\overline{B}_{}(m_K^2,m_\eta^2,u)\,( 1/32\,m_\eta^2 - 1/32\,u + 3/64\,\Delta 
+ 3/64\,\Sigma )
\nonumber\\&&
+\overline{B}_{1}(m_\pi^2,m_K^2,u)\,(1/16\Delta)\nonumber\\&&
+\overline{B}_{1}(m_K^2,m_\eta^2,u)\,( 1/16\,m_\eta^2 - 1/32\,\Delta 
- 1/32\,\Sigma )\,,
\\
F_\pi^4\, G_6(s,t,u) &=&F_\pi^2(-1/2\,s\,+1/2\Sigma)\nonumber\\&&
+(16\pi^2)^{-1}\,( 3/32\,m_\eta^2\,s-1/32\,m_\eta^2\,(t-u)-1/16\,m_\eta^2\,\Sigma 
\nonumber\\&&\hspace{2cm}- 1/64\,s\,\Delta +31/192\,s\,\Sigma + 1/16\,s^2 
- 7/192\,(t-u)\,\Delta 
\nonumber\\&&\hspace{2cm}+ 1/64\,(t-u)\,\Sigma - 1/48\,(t-u)^2 
+ 1/96\,\Delta\,\Sigma 
- 13/96\,\Sigma^2 )\nonumber\\&&
+L_1^r\,\,(-4\,s\,(t-u)+2\,s^2+2(t-u)^2-8\Delta^2)\nonumber\\&&
+L_2^r\,\,(2\,s\,(t-u)-8\,s\,\Sigma+5\,s^2+(t-u)^2+4\Sigma^2)\nonumber\\&&
+L_3^r\,\,(\,s^2+(t-u)^2-2\Delta^2)\nonumber\\&&
+L_4^r\,\,(-4\,s\,\Sigma+4(t-u)\Sigma+8\Delta^2)\nonumber\\&&
+L_5^r\,\,(2\,s\,\Delta-2\,s\,\Sigma-2\Delta\Sigma+2\Delta^2)\nonumber\\&&
+L_6^r\,\,(-8\Delta^2+8\Sigma^2)\nonumber\\&&
+L_8^r\,\,(-4\Delta^2+4\Sigma^2)\nonumber\\&&
+\overline{A}(m_\pi^2)\,  - 21/32\,s - 5/96\,(t-u) - 7/32\,\Delta 
+ 13/24\,\Sigma )\nonumber\\&&
+\overline{A}(m_K^2)\,( 1/96\,m_\eta^2 - 3/16\,s + 1/48\,(t-u) + 31/576\,\Delta 
+ 5/64\,\Sigma )\nonumber\\&&
+\overline{A}(m_\eta^2)\,( 9/160\,m_\eta^2 + 3/32\,s + 1/32\,(t-u) + 7/64\,\Delta 
- 49/320\,\Sigma)\,.\nonumber\\
\end{eqnarray}

The finite part of the one-loop integrals are denoted 
by $\overline{A}, \overline{B}_j$ as defined in \cite{ABT1}.

\subsection{Results at order $p^6$}

The full result at order $p^6$ is rather cumbersome\footnote{It can be obtained from the
website \cite{formulas} or from the authors upon request.}. 
In this section we quote
only the dependence on the order $p^6$ constants. This contribution
can be written exactly in the form of the subthreshold expansion 
(\ref{defci})
with as nonzero combinations, normalized to $F_\pi^4\,m_{\pi^+}^{2i+2j}$ and
$F_\pi^4\,m_{\pi^+}^{2i+2j+1}$ for $c^+_{ij}$ and $c^-_{ij}$ respectively,
\ba
c^+_{00} &=&
 16\,m_{K^+}^2\,m_{\pi^+}^4 \, ( C^r_{1} + 4\,C^r_{2} + C^r_{5} + C^r_{6}
+ 2\,C^r_{7} - C^r_{12} - 4\,C^r_{13} - 2\,C^r_{14}
- 3\,C^r_{15}
\nonumber\\ 
         & &\qquad\qquad\qquad
 - 4\,C^r_{16}
+ 3\,C^r_{19} + 6\,C^r_{20} + 12\,C^r_{21} 
         - C^r_{26} - 4\,C^r_{28} + 3\,C^r_{31} + 6\,C^r_{32} )
\nonumber\\
    & &+ 16\,m_{K^+}^4\,m_{\pi^+}^2 \, ( C^r_{1} + 4\,C^r_{2} + 2\,C^r_{6} 
+ 2\,C^r_{7} + C^r_{8} - C^r_{12} - 4\,C^r_{13} 
- 4\,C^r_{15} - 4\,C^r_{16}
\nonumber\\
    & &\qquad\qquad\qquad- 2\,C^r_{17} + 3\,C^r_{19} 
+ 8\,C^r_{20} + 24\,C^r_{21} - C^r_{26} - 4\,C^r_{28} +
         3\,C^r_{31} + 8\,C^r_{32} )\,,
\nonumber\\
c^+_{10} &=& 8\,m_{K^+}^2\,m_{\pi^+}^2 \, (  - 4\,C^r_{1} - 16\,C^r_{2} 
+ 2\,C^r_{4} - C^r_{5} - 3\,C^r_{6} - 4\,C^r_{7} - C^r_{8}
+ 4\,C^r_{12} + 10\,C^r_{13} 
\nonumber\\
& &\qquad\qquad\qquad+ C^r_{14} + 5\,C^r_{15} + 2\,C^r_{17} 
- 2\,C^r_{22} - 2\,C^r_{25} - C^r_{26}- 2\,C^r_{29} - 4\,C^r_{30} )
\nonumber\\
      & &+ 8\,m_{K^+}^4 \, (  - C^r_{1} - 4\,C^r_{2} - 2\,C^r_{6} - 2\,C^r_{7} 
- C^r_{8} - 2\,C^r_{12} - 4\,C^r_{13} 
\nonumber\\
& &\qquad\qquad -C^r_{14} + 4\,C^r_{16} + C^r_{26} + 4\,C^r_{28} )
\nonumber\\
      & &+ 8\,m_{\pi^+}^4 \, (  - C^r_{1} - 4\,C^r_{2} - C^r_{5} - C^r_{6} 
- 2\,C^r_{7} - 2\,C^r_{12} 
\nonumber\\
& &\qquad\qquad- 2\,C^r_{13} +C^r_{14} + 1/2\,C^r_{15} + 4\,C^r_{16} - C^r_{17} + C^r_{26} 
+ 4\,C^r_{28} )\,,
\nonumber\\
c^-_{00} &=&     16\,m_{K^+}\,m_{\pi^+}^4 \, ( C^r_{15} + 2\,C^r_{17} )
\nonumber\\
       & &+ 16\,m_{K^+}^3\,m_{\pi^+}^3 \, ( 4\,C^r_{4} + 2\,C^r_{14} + 2\,C^r_{15} - 4\,C^r_{22} + 4\,C^r_{25}
+ 2\,C^r_{26}- 4\,C^r_{29} )\,,
\nonumber\\
c^+_{20} &=& m_{K^+}^2 \, ( 12\,C^r_{1} + 48\,C^r_{2} - 8\,C^r_{4} + C^r_{5} 
+ 10\,C^r_{6} + 8\,C^r_{7} + 4\,C^r_{8}
          + C^r_{10} 
\nonumber\\
& &\qquad\qquad+ 4\,C^r_{11} - 2\,C^r_{12} - 4\,C^r_{13} + 2\,C^r_{22} 
- 4\,C^r_{23} + 4\,C^r_{25} )
\nonumber\\
       & &+ m_{\pi^+}^2 \, ( 12\,C^r_{1} + 48\,C^r_{2} - 8\,C^r_{4} + 4\,C^r_{5} 
+ 5\,C^r_{6} + 8\,C^r_{7} + C^r_{8} +
         C^r_{10} 
\nonumber\\
& &\qquad\qquad+ 2\,C^r_{11} - 2\,C^r_{12} - 10\,C^r_{13} + 2\,C^r_{22} 
- 4\,C^r_{23} + 4\,C^r_{25} )\,,
\nonumber\\
c^-_{10} &=& 8\,m_{K^+}\,m_{\pi^+}^2 \, (  - 4\,C^r_{4} - C^r_{6} - C^r_{8} 
+ C^r_{10} 
+ 2\,C^r_{11} - 2\,C^r_{12} - 6\,
         C^r_{13} + 2\,C^r_{22} - 2\,C^r_{25} )
\nonumber\\
      & & + 8\,m_{K^+}^3\,m_{\pi^+} \, (  - 4\,C^r_{4} - C^r_{5} - 2\,C^r_{6} 
      + C^r_{10} 
      + 4\,C^r_{11} - 2\,C^r_{12} - 12
         \,C^r_{13} + 2\,C^r_{22}
\nonumber\\& &\qquad\qquad\qquad
 - 2\,C^r_{25} )\,,
\nonumber\\
c^+_{01} &=&16\,m_{K^+}^2\,m_{\pi^+}^2 \, ( C^r_{6} + C^r_{8} + C^r_{10} + 2\,C^r_{11} 
- 2\,C^r_{12} 
- 2\,C^r_{13} + 2\,C^r_{22} + 4\,C^r_{23} )
\nonumber\\
        & &+16\,m_{K^+}^4\,m_{\pi^+}^2 \, ( C^r_{5} + 2\,C^r_{6} + C^r_{10} 
+ 4\,C^r_{11} 
- 2\,C^r_{12} 
- 4\,C^r_{13} + 2\,C^r_{22} + 4\,C^r_{23} )\,,
\nonumber\\
c^+_{30} &=&1/2\,(  - 7\,C^r_{1} - 32\,C^r_{2} + 2\,C^r_{3} + 10\,C^r_{4} )\,,
\nonumber\\
c^-_{20} &=&6\,m_{K^+} \, (  - C^r_{1} + 2\,C^r_{3} + 2\,C^r_{4} )\,,
\nonumber\\
c^+_{11} &=&8\,m_{K^+}^2 \, ( 3\,C^r_{1} + 6\,C^r_{3} - 2\,C^r_{4} )\,,
\nonumber\\
c^-_{01} &=&32\,m_{K^+}^3 \, (  - C^r_{1} + 2\,C^r_{3} + 2\,C^r_{4} )\,.
\ea

Notice that the combinations $ (  - C^r_{1} + 2\,C^r_{3} + 2\,C^r_{4} )$
shows up in both $c^-_{20}$ and $c^-_{01}$. 

\section{Resonance estimate of the contribution from the $p^6$ constants} 


Up to now we relied only on chiral symmetries to calculate the amplitude 
function at low energy. 
In order to give an estimate of the $p^6$ LECs,
we assume our process to be saturated by the exchange of vector and scalar
meson
resonances. The general formalism of resonance saturation (RS) in ChPT was
described 
in \cite{Ecker1}, \cite{Ecker2}. 
The places where comparisons
with experiment are available are in general in reasonable agreement with the
estimates obtained via RS. 

For both types of exchange, we only consider the polynomial 
contributions to
$K\pi$-scattering starting at $\mathcal{O}(p^6)$, thus directly corresponding
to the $C_i^r$'s LEC's contribution.

The vector resonances are included
through the matrix of fields
$V^\mu$ \cite{ABT3} with Lagrangian
\be
{\cal L}_V = -\frac{1}{4}\langle V_{\mu\nu}V^{\mu\nu}\rangle
+\frac{1}{2}m_V^2\langle V_\mu V^\mu\rangle
-\frac{ig_V}{2\sqrt{2}}\langle V_{\mu\nu}[u^\mu,u^\nu]\rangle
+f_\chi\langle V_\mu[u^\mu,\chi_-]\rangle
\label{vector}
\ee
while for the scalar meson nonet, the matrix of fields $S$, we consider
\be
{\cal L}_S = \frac{1}{2} \langle \nabla^\mu S \nabla_\mu S 
 - M^2_S S^2 \rangle  
 + c_d \langle Su^\mu u_\mu \rangle + c_m \langle S \chi_+ \rangle 
\label{scalar}
\ee
After integration of the resonance fields, the Lagrangians relevant to the
present case read 
\begin{equation}
{\cal L}_V = -\frac{1}{4 M^2_V} \left\langle \left( i g_V\,
\nabla_\mu [ u^\nu,u^\mu ] - f_\chi \sqrt{2}\, [ u^\nu,\chi_- ] \right)^2 
\right\rangle
\label{LagInt}
\end{equation}

\begin{equation}
\label{LagInt2}
{\cal L}_S = \frac{1}{2 M^4_S} \left\langle 
\left( c_d \nabla^\nu ( u_\mu u^\mu )
 + c_m \nabla^\nu \chi_+ 
 \right)^2 \right\rangle
\end{equation}
where we use \cite{BCEGS2}
\begin{eqnarray}
 f_\chi = -0.025,\quad  g_V = 0.09,\quad 
 c_m = 42 \mbox{ MeV},\quad
  c_d = 32 \mbox{ MeV}, 
\end{eqnarray}
and the masses are
\begin{eqnarray}
m_V = m_\rho = 0.77 \mbox{ GeV}, & m_A = m_{a_1} = 1.23 \mbox{ GeV}, &
m_S = 0.98 \mbox{ GeV}.
\end{eqnarray}
The numerical results from both contributions
to the subthreshold expansion parameters
are listed in Table~\ref{tab:subthres}. The contribution to the full amplitude
at order $p^6$ corresponds precisely to the expansion (\ref{defci})
including only these
subthreshold constants.
The explicit expressions for the nonzero constants are
\begin{eqnarray}
c^+_{00}&=& -16\, f_\chi\,g_V/(M_V^{2}\sqrt{2}) \, (
          m_K^2\,m_\pi^4
          + m_K^4\,m_\pi^2
          )
       + 4\,c_m^2/M_S^{4} \, (
          m_K^2\,m_\pi^4
          + m_K^4\,m_\pi^2
          )\nonumber\\&&
       - 16\, f_\chi^2/M_V^{2} \, (
           m_K^2\,m_\pi^4
          + m_K^4\,m_\pi^2
          )
       -2\, g_V^2/M_V^{2} \, (
          m_K^2\,m_\pi^4
          + m_K^4\,m_\pi^2
          )\,,\nonumber\\
c^+_{10} &=&   8\, f_\chi\,g_V/(M_V^{2}\sqrt{2}) \, (
          2\,m_K^2\,m_\pi^2
          + \,m_K^4
          + \,m_\pi^4
          )
       - c_d\,c_m/M_S^{4} \, (
          12\,m_K^2\,m_\pi^2
          ) \nonumber\\&&
       + c_d^2/M_S^{4} \, (
           4\,m_K^2\,m_\pi^2
          )
      + 2\, c_m^2/M_S^{4} \, (
           3\,m_K^2\,m_\pi^2
          - \,m_K^4
          - \,m_\pi^4
          )\nonumber\\&&
      + 8\, f_\chi^2/M_V^{2} \, (
           \,m_K^2\,m_\pi^2
          + \,m_K^4
          + \,m_\pi^4
          )
       + g_V^2/M_V^{2} \, (
           3\,m_K^2\,m_\pi^2
          + m_K^4
          + m_\pi^4
          )\,,\nonumber\\
c^-_{00}&=& c_d\,c_m/M_S^{4} \, (
           16\,m_K^3\,m_\pi^2
          )
       + c_m^2/M_S^{4} \, (
           8\,m_K^3\,m_\pi^2
          )\nonumber\\&&
       + f_\chi^2/M_V^{2} \, (
           96\,m_K^3\,m_\pi^2
          )
       - g_V^2/M_V^{2} \, (
           4\,m_K^3\,m_\pi^2
          )\,,\nonumber\\
c^+_{20}&=& -3\,  f_\chi\,g_V/(M_V^{2}\sqrt{2}) \, (
           \,m_K^2
          + \,m_\pi^2
          )
       + 5/2\,c_d\,c_m/M_S^{4} \, (
           m_K^2
          + m_\pi^2
          )\nonumber\\&&
       -7/4\, c_d^2/M_S^{4} \, (
           m_K^2
          + m_\pi^2
          )
       - 7/8\,g_V^2/M_V^{2} \, (
           m_K^2
          + m_\pi^2
          )\,,\nonumber\\
c^-_{10}&=& 24\, f_\chi\,g_V/(M_V^{2}\sqrt{2}) \, (
           m_K\,m_\pi^2
          + m_K^3
          )
       -4\, c_d\,c_m/M_S^{4} \, (
          m_K\,m_\pi^2
          + m_K^3
          )\nonumber\\&&
       -2\, c_d^2/M_S^{4} \, (
          m_K\,m_\pi^2
          + m_K^3
          )
       +3\,g_V^2/M_V^{2} \, (
          m_K\,m_\pi^2
          + m_K^3
          )\,,\nonumber\\
c^+_{01}&=& 16\, f_\chi\,g_V/(M_V^{2}\sqrt{2}) \, (
           m_K^2\,m_\pi^2
          + m_K^4
          )
       + 8\,c_d\,c_m/M_S^{4} \, (
           m_K^2\,m_\pi^2
          + m_K^4
          )\nonumber\\&&
       + 4\,c_d^2/M_S^{4} \, (
           m_K^2\,m_\pi^2
          + m_K^4
          )
       + 2\,g_V^2/M_V^{2} \, (
          m_K^2\,m_\pi^2
          + m_K^4
          )\,,\nonumber\\
c^+_{30}&=&   c_d^2/M_S^{4} \, (
           7/8
          )
          + g_V^2/M_V^{2} \, (
           3/16
          )\,,\nonumber\\
c^-_{20}&=& c_d^2/M_S^{4} \, (
          3/2\,m_K
          )
       + g_V^2/M_V^{2} \, (
           3/4\,m_K
          )\,,\nonumber\\
c^+_{11}&=& -6\, c_d^2/M_S^{4} \, (
           m_K^2
          )
       + g_V^2/M_V^{2} \, (
           m_K^2
          )\,,\nonumber\\
c^-_{01}&=& c_d^2/M_S^{4} \, (
           8\,m_K^3
          )
       + g_V^2/M_V^{2} \, (
           4\,m_K^3
          )\,.
\end{eqnarray}

\TABLE{
\begin{tabular}{lrrrcrrr}
\hline
           & Vector & Scalar & Sum Reso& chiral order& $p^2$ & $p^4$ & $p^6$\\
\hline
$c^+_{00}$ &$-$0.02  &  0.13   &  0.11    &2& 0     &0.122    &0.007    \\
$c^+_{10}$ &  0.018  &$-$0.063 & $-$0.045 &2& 0.5704&$-$0.113 &0.460    \\
$c^-_{00}$ &  0.21   &  0.17   &  0.38    &2& 8.070 &0.311    &0.017    \\
$c^+_{20}$ &$-$0.0053&  0.0023 & $-$0.0030&4& ---   &0.0256   &$-$0.0254\\
$c^-_{10}$ & $-$0.11 &$-$0.04  & $-$0.15  &4& ---   &$-$0.0254&0.121    \\
$c^+_{01}$ &$-$0.27  &  0.28   &  0.01    &4& ---   &1.667    &1.492    \\
$c^+_{30}$ &  0.00026&  0.00010&  0.00036 &6& ---   &0.00121  &0.00071  \\
$c^-_{20}$ &  0.0037 &  0.00060&  0.0043  &6& ---   &0.00478  &0.00320  \\
$c^+_{11}$ &  0.017  &$-$0.008 &  0.009   &6& ---   &$-$0.126 &$-$0.006 \\
$c^-_{01}$ &  0.25   &  0.04   &  0.29    &6& ---   &0.229    &0.196    \\ 
\hline
\end{tabular}
\caption{\label{tab:subthres} The resonance contributions to the 
subthreshold parameters. The units are $m_{\pi^+}^{2i+2j}$ and
$m_{\pi^+}^{2i+2j+1}$ for $c^+_{ij}$ and $c^-_{ij}$ respectively.
We have also shown the chiral order at which they first have tree level
contributions as well as the contributions with the $L_i^r=C_i^r=0$
at $\mu=0.77$~GeV.}
}

Even if many of our results only get a small contribution from the above RS
arguments, these
estimates are in general a major source of uncertainty in the $\mathcal{O}
(p^6)$ terms.
The estimates from resonance exchange for the masses and decay constants
and the related sigma terms are the most uncertain because of the
simple treatment of the scalar sector.
These are discussed in more detail in \cite{ABT1} and \cite{Pich}. 
Here we have set many
effects, e.g. the $d_m$ term of \cite{ABT1},
equal to zero, the naive
size estimate of \cite{ABT1} led to anomalously large NNLO corrections.
The estimates of the $K_{\ell4}$ amplitudes can be found in
\cite{ABT3} after the work of \cite{BCG}. The effect of varying
these was studied in \cite{ABT3} and found to be reasonable.

The above procedure is obviously subtraction
point dependent and is normally only performed to leading order in the
expansion in $1/N_c$, with
$N_c$ the number of colours. Many other approaches
exist, some recent relevant papers
addressing this issue are \cite{Pich,BGLP,MHA} and
references therein. A systematic study of this issue is clearly
important, for our present first study the estimates are sufficient.

\section{A first numerical look}
\label{firstnum}

In this section we present a first look at the numerical results
for the two loop amplitudes. We choose as input the pion decay constant,
the charged pion mass, the charged kaon mass and the physical eta mass.
\ba
F_\pi &=& 92.4~\mbox{MeV}\,,\qquad m_\pi = m_{\pi^+} = 139.56995~\mbox{MeV}\,,
\nonumber\\
m_K &=& m_{K^+} = 493.677~\mbox{MeV}  \,,\quad m_\eta = 547.3~\mbox{MeV}\,.
\ea
The subtraction scale $\mu = 770~\mbox{MeV}$ is used throughout
the paper unless otherwise mentioned explicitly.

We present the results for the subthreshold parameters with all $L_i^r$
and $C_i^r$ set equal to zero at the scale of the rho mass. Notice that
the very different sizes of the various quantities are to a large extent
given by their normalization in powers of $m_K$
and $m_\pi$.

The order $p^4$ results differ somewhat from those quoted in \cite{BKM}
and \cite{AB}. The reason for this is that some variation in the choice
of precisely what is called $p^4$ and $p^6$ is possible. The choice we
have made is different from those in the mentioned papers.
Especially $c^+_{00}$ suffers from this numerically.
E.g. taking the eta mass given by
the Gell-Mann-Okubo formula instead, the numerical $p^4$ result for it changes
to $0.16$. We conclude that we are also in numerical
agreement with those papers.

The higher order corrections look very large when looked upon as the
contributions to the various terms in the expansions. This is partly due
to cancellations making some quantities very sensitive
to higher order corrections.

\FIGURE{
\includegraphics[width=7cm,angle=270]{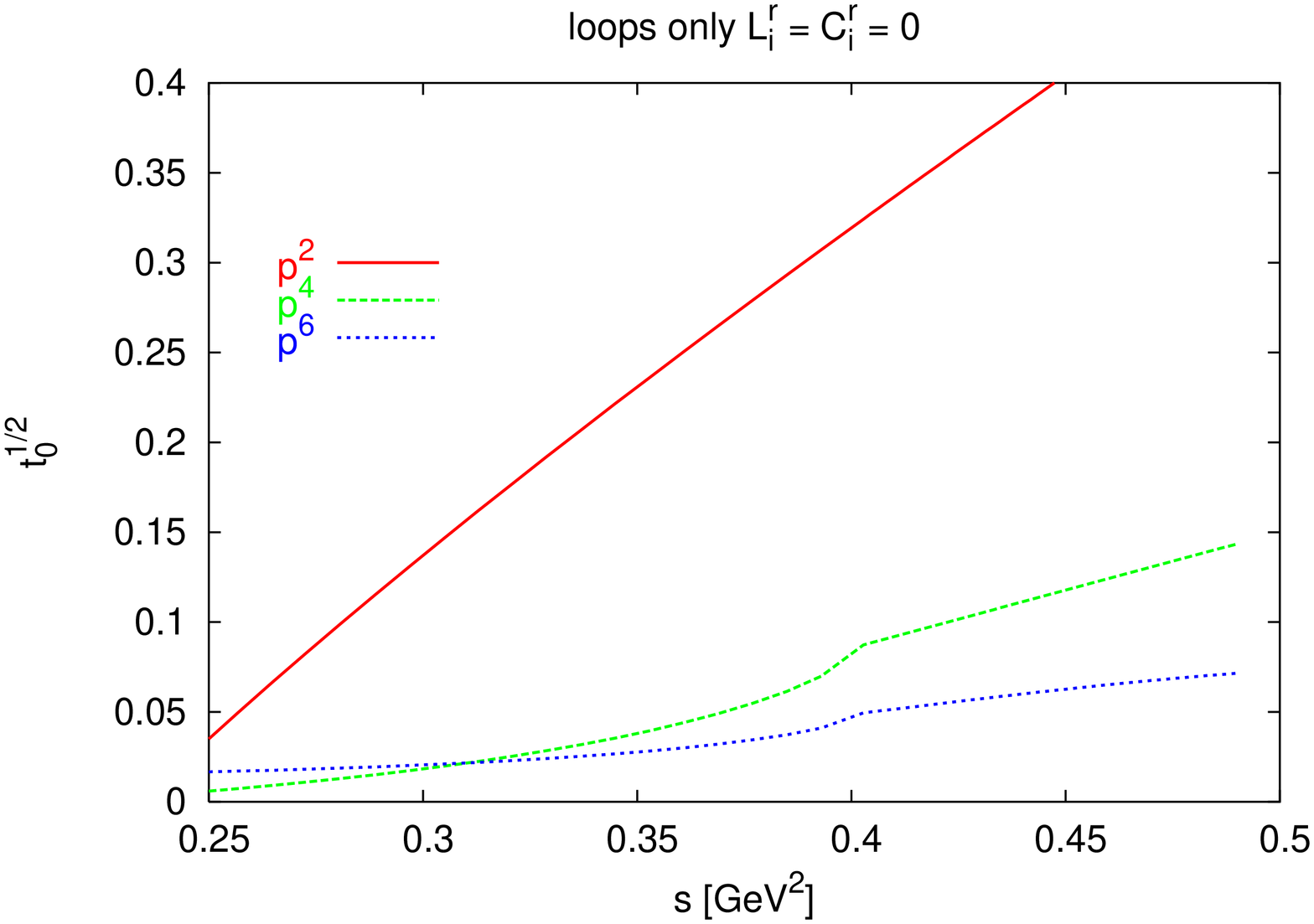}
\caption{\label{fig:t120} The corrections at order $p^2$, $p^4$ and $p^6$
to the $S$ partial wave in the $I=1/2$ channel with all LECs
set to zero at $\mu=0.77$~GeV.}
}
\FIGURE{
\includegraphics[width=7cm,angle=270]{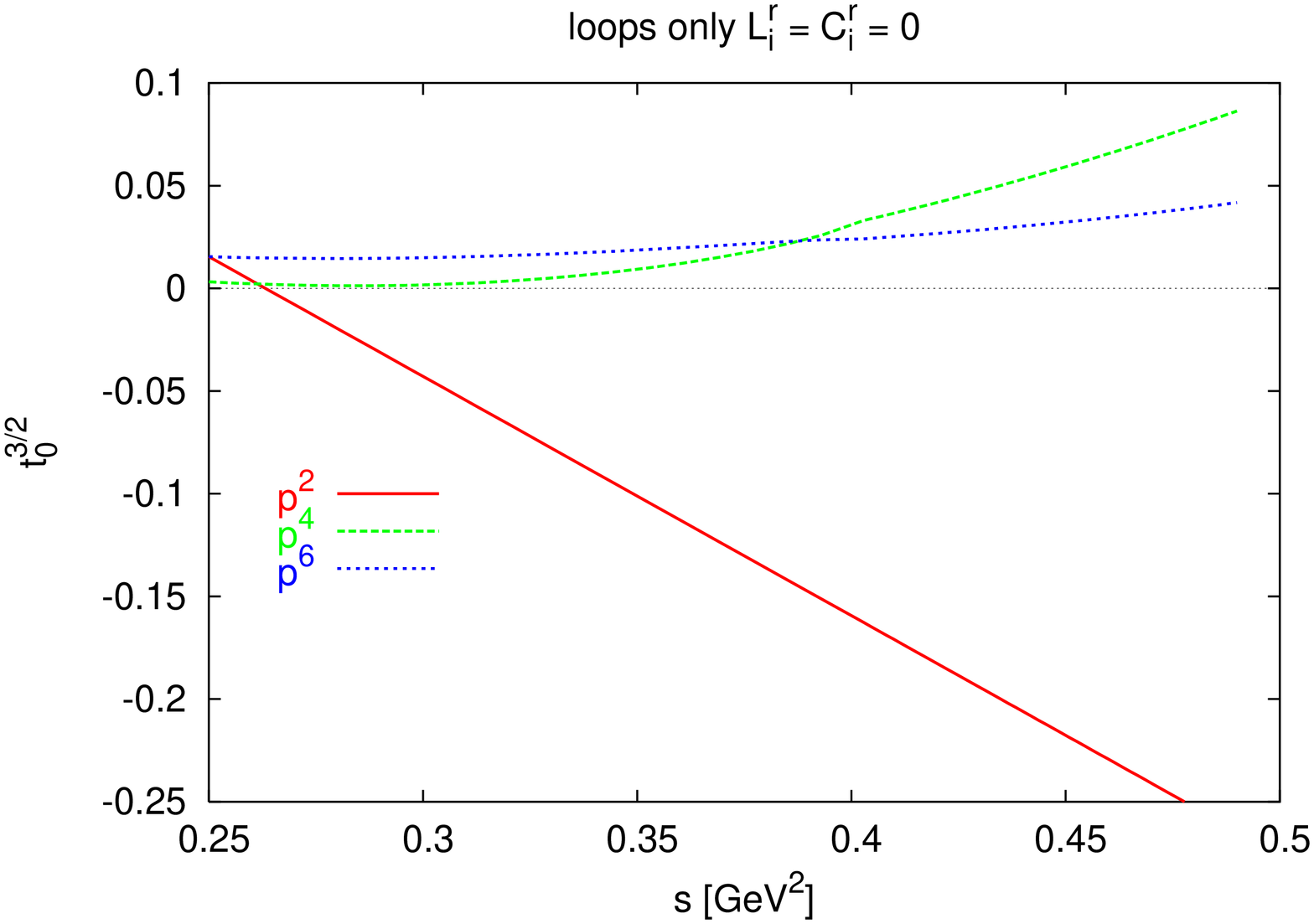}
\caption{\label{fig:t320} The corrections at order $p^2$, $p^4$ and $p^6$
to the $S$ partial wave in the $I=3/2$ channel with all LECs
set to zero at $\mu=0.77$~GeV.}
}
We also present some plots of the $S$ channel partial waves both in the
$I=1/2$ and $I=3/2$ channel. These are shown in respectively
Fig.~\ref{fig:t120}
and Fig.~\ref{fig:t320}. Two points of interest
are $s=0.263$~GeV$^2$ for the subthreshold expansion
and the threshold at $s=0.401$~GeV$^2$. From the sizes of the
contributions at orders $p^2$, $p^4$ and $p^6$ as shown it is obvious
that there seems to be a better convergence near threshold than at
the subthreshold point.
Contrary to the $\pi\pi$ scattering case, the lowest order result has already
nonlinearities. The amplitude at this level is perfectly linear in
$s$, $t$ and $u$ but in taking the partial waves, $t$ and $u$ depend
nonlinearly on $s$ via $q^2_{\pi K}$.
The partial waves have been extracted from the amplitudes using a five point
Gaussian integration over $\cos\theta$.

\TABLE{
\begin{tabular}{rrrrr}
\hline
  & $p^2$ & $p^4$ & $p^6$ & Reso \\
\hline
$     a^{1/2}_0$ &    0.142 &    0.035 &    0.022 &    0.013\\
$10   a^{1/2}_1$ &    0.100 &    0.006 &    0.056 & $-$0.010\\  
$10^3 a^{1/2}_2$ &    0     &    0.142 &    0.029 & $-$0.029\\
$10   a^{3/2}_0$ & $-$0.708 &    0.145 &    0.105 & $-$0.048\\
$10^2 a^{3/2}_1$ &    0     &    0.003 &    0.308 & $-$0.018\\  
$10^3 a^{3/2}_2$ &    0     &    0.092 & $-$0.139 &    0.002\\
$10   b^{1/2}_0$ &    0.664 &    0.311 &    0.112 &    0.191\\
$10^2 b^{1/2}_1$ & $-$0.141 &    0.001 &    0.165 &    0.007\\  
$10^3 b^{1/2}_2$ &    0     & $-$0.065 & $-$0.174 &    0.041\\
$10   b^{3/2}_0$ & $-$0.482 &    0.191 &    0.052 & $-$0.087\\
$10^3 b^{3/2}_1$ &    0     & $-$0.204 &    0.542 &    0.206\\  
$10^3 b^{3/2}_2$ &    0     & $-$0.074 &    0.163 & $-$0.028\\
$T^+_{CD}$       & 1.141    & 0.010    &    0.831 &    0.011\\
\hline
\end{tabular}
\caption{\label{tab:aij1} The contributions at order $p^2$, $p^4$ and $p^6$
for the scattering lengths and ranges 
and the amplitude at the Cheng-Dashen point with the LECs set equal to zero
as well as the resonance estimate to
order $p^6$. }
}
In order to show the convergence also around threshold we present as well
in Table~\ref{tab:aij1} the contributions at order $p^2$, $p^4$ and $p^6$
the scattering lengths and ranges and the value of the $T^+$
amplitude at the Cheng-Dashen point
\be
s=u= m_K^2\,,\quad t=2m_\pi^2\,,
\ee
as well.
These will also be studied in more detail
later on when we add the contributions from the LECs and compare to
experimental and dispersion relation results.

\section{Input parameters}
\label{inputs}

For our $SU(3)$ ChPT results we use as inputs the masses and decay constants
given in Sect.~\ref{firstnum} and a subtraction constant $\mu = 770$~MeV.
We work in the limit of exact isospin.

In addition we use the full refit of 
the $L_i^r$ $(i=1,2,3,5,7,8)$ to order $p^6$ using a range
of values for $L_4^r$ and $L_6^r$ as input. These
values of $L_i^r,i=1,\ldots,8$, are used  to evaluate the matrix element.
This is the same procedure
used in \cite{BD} and \cite{BDT}
to study the variation of some observables as a function of
the vector ($L_4^r$,$L_6^r$). 
The experimental inputs used for the fitting procedure are: {\it i)} 
the values of the
$K_{\ell4}$ form-factors as measured
by the E865 experiment \cite{Pislak1,Pislak2}, $f_s(0),g_p(0),f_s^\prime(0),
g_p^\prime(0)$, {\it ii)} the pseudoscalar 
decay constants $F_{\pi^\pm}, F_{K^\pm}$ and {\it iii)} the 
masses of the pseudoscalar mesons, $m_{\pi^\pm},m_{K^\pm},m_{\eta}$. 
The performed fits correspond to fit 10 in \cite{ABT4}
but with different input values for the vector $(L_4^r,L_6^r)$. These
represent the only free parameters, in the analysis. The quark-mass ratio
$m_s/\hat{m}$ used is fixed to be $24$. The variation
inside the range $(20-30)$ was studied in \cite{ABT4}.

The estimates of the $p^6$ contributions to $\pi K$ scattering we use
are those given above. These lead to the central value
contributions to the various threshold parameters given in
Table \ref{tab:aij1}. The uncertainty on these is quite considerable.
Other resonance estimates of $\pi K$ scattering can be
found in \cite{BKM3,JOP}.

The $\pi K$ scattering amplitude also obeys relations from crossing and
unitarity. A new recent analysis using these  methods is \cite{BDM}.
Once the choice of the high energy input is done, no more freedom is allowed.
The constraints at the matching point are stronger than in the case
of the Roy analysis of $\pi\pi$ scattering. We also quote for comparison
the results from the older analysis of \cite{Lang}. This is what we
use as our main ``experimental'' input for $\pi K$ scattering.

\section{Numerical Analysis}
\label{numerical}

\subsection{$\pi K$ only}

\TABLE{
\begin{tabular}{rrrr}
\hline
           & Fit 10 & \cite{BDM} & \cite{Lang}\\
\hline
$c^+_{00}$ &$   0.278$&$ 2.01\pm1.10$ & $-0.52\pm2.03$\\
$c^+_{10}$ &$   0.898$&$ 0.87\pm0.08$ & $ 0.55\pm0.07$\\
$c^-_{00}$ &$   8.99$&$ 8.92\pm0.38$ & $ 7.31\pm0.90$\\
$c^+_{20}$ &$ 0.003$&$ 0.024\pm0.006$ & \\
$c^-_{10}$ &$ 0.088$&$ 0.31\pm0.01$ & $0.21\pm0.04 $\\
$c^+_{01}$ &$ 3.8$&$ 2.07\pm0.10$ & $2.06\pm0.22 $\\
$c^+_{30}$ &$ 0.0025$&$ 0.0034\pm0.0008$ & \\
$c^-_{20}$ &$ 0.013$&$ 0.0085\pm0.0001$ & \\  
$c^+_{11}$ &$ -0.10$&$-0.066\pm0.010 $ & \\
$c^-_{01}$ &$ 0.71$&$0.62\pm0.06 $ & $0.51\pm0.10 $\\
$c^+_{02}$ &$ 0.23$&$0.34\pm0.03 $ & \\ 
\hline
\end{tabular}
\caption{\label{tab:subthres2} The 
subthreshold parameters. The units are $m_{\pi^+}^{2i+2j}$ and
$m_{\pi^+}^{2i+2j+1}$ for $c^+_{ij}$ and $c^-_{ij}$ respectively.
Shown are the results for fit 10 of \cite{ABT4} and the dispersive
results from \cite{BDM} and \cite{Lang}.}
}
Let us first look at the subthreshold expansions and compare the
dispersive calculations with our results. The results from the two
analyses \cite{Lang} and \cite{BDM} can be found in Table~\ref{tab:subthres2}
together with our calculation for the $L_i^r$ corresponding
to fit 10 of \cite{ABT4}.

The results for
$c^+_{10}$ and $c^-_{00}$, which are the only two that obtain a lowest
order contribution, are shown in Figs.~\ref{fig:cp10}(a) and \ref{fig:cm00}(b).
$c^-_{00}$ has a large lowest order contribution and shows a reasonable
convergence over the entire region of variation of $(L_4^r,L_6^r)$ we have
covered.
It is also in good agreement with the
dispersive calculation inside the whole region.
The result for $c^+_{10}$ shows somewhat less good convergence but it is still
acceptable. This is what was used in earlier analyses of $\pi K$ scattering to
get a determination of the $1/N_c$ suppressed constant $L_4^r$.
As can be seen in Fig.~\ref{fig:cp10}(a) the matching with the dispersive
values is obtained within the region $ L_4^r \approx -0.1\,10^{-3}$
with only a fairly weak dependence on the value of $L_6^r$.
\FIGURE{
\begin{minipage}{7.001cm}
\includegraphics[width=7cm,angle=270]{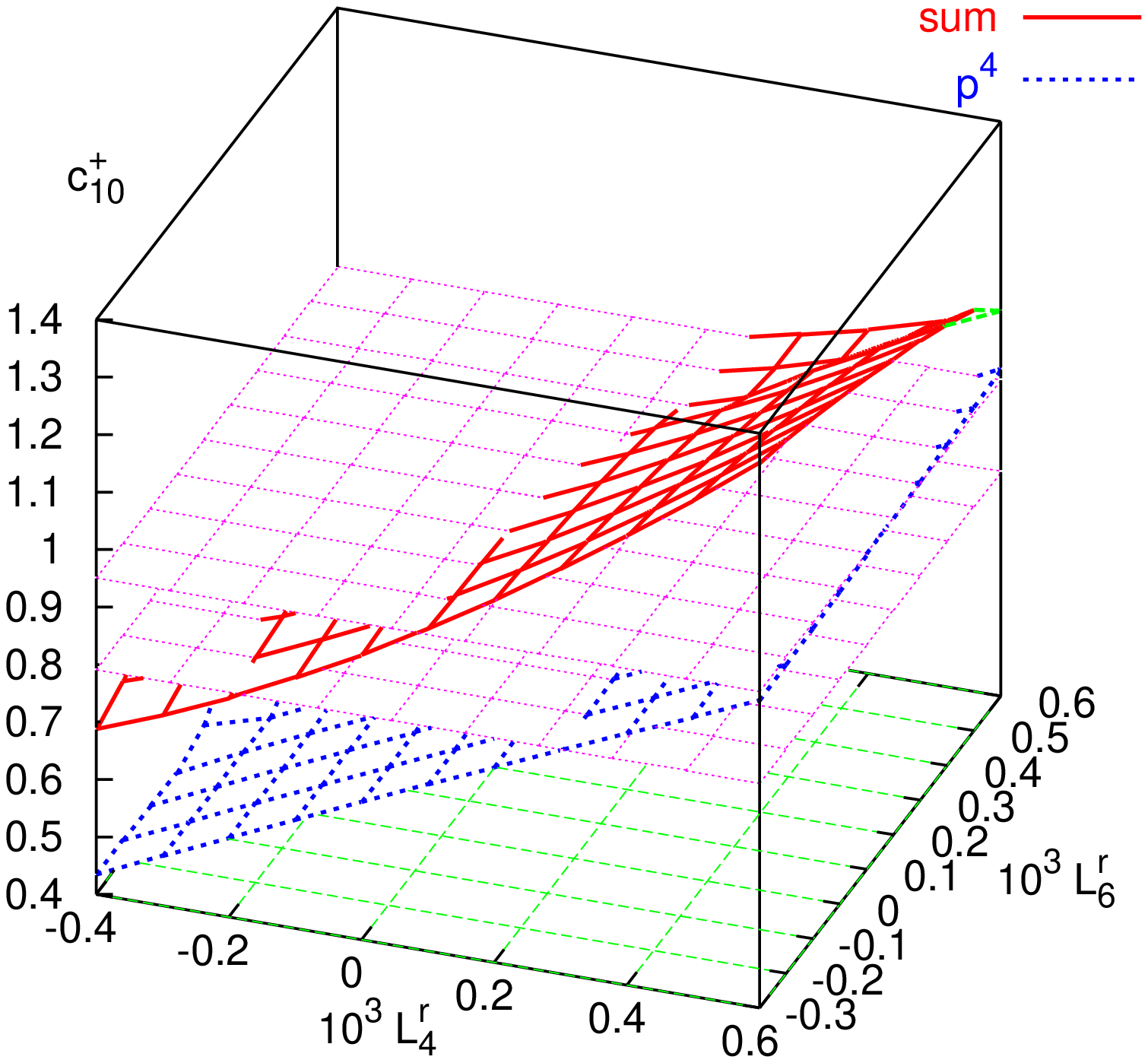}
\centerline{(a)}
\end{minipage}
\begin{minipage}{7.001cm}
\includegraphics[width=7cm,angle=270]{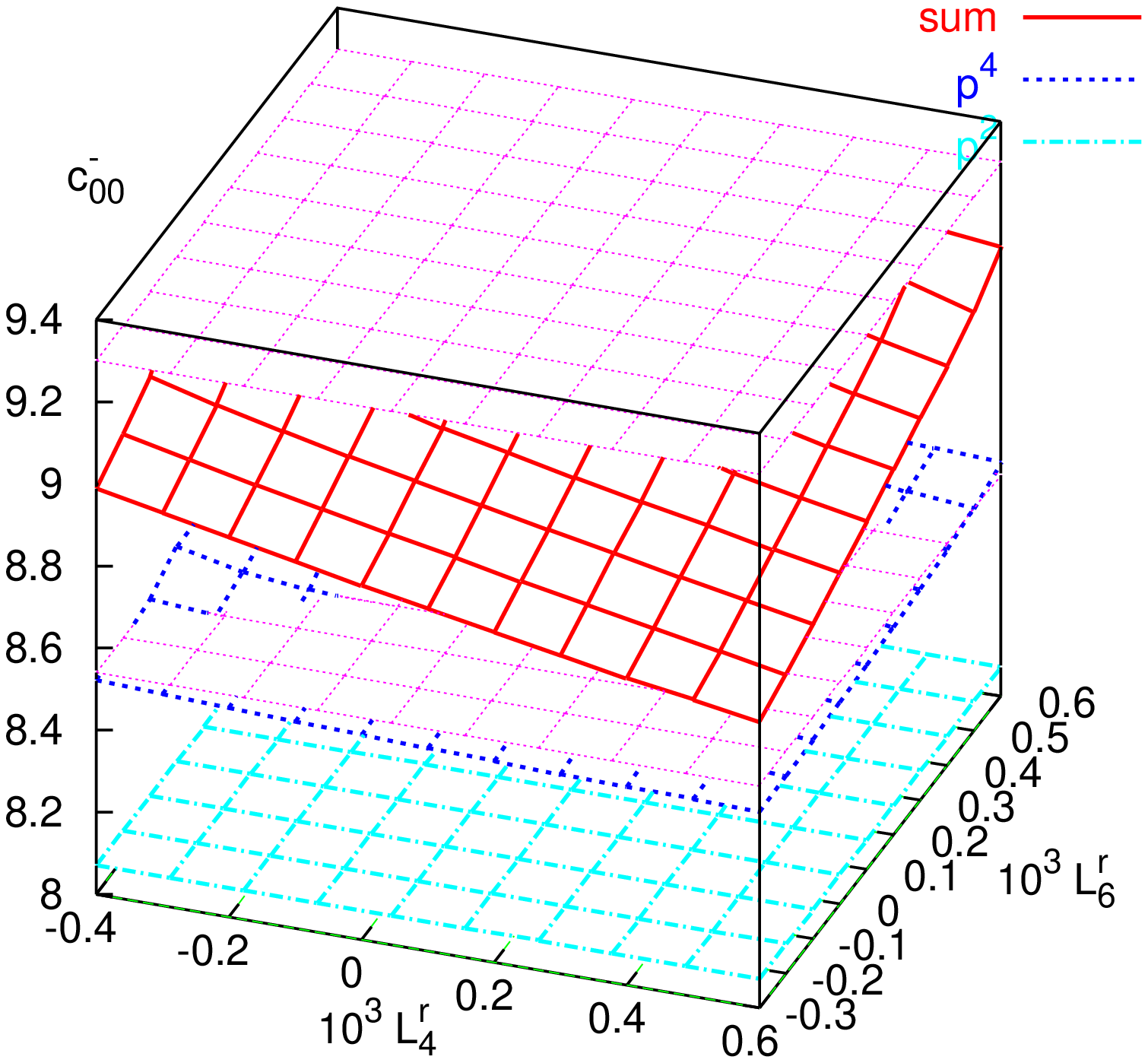}
\centerline{(b)}
\end{minipage}
\caption{\label{fig:cp10}\label{fig:cm00} (a) The subthreshold
parameter $c^+_{10}$ as a function
of the input values of $L_4^r$ and $L_6^r$ used in the $L_i^r$ determination.
The unlabeled planes are the result of \cite{BDM} with their errors.
  (b) The same for the subthreshold parameter $c^-_{00}$.}
}

The results for the other subthreshold parameters are more difficult to
interpret. They show a variety of behaviours:{\it i)} some 
subthreshold parameters display indication
of reasonable convergence while others obviously do not converge.
{\it ii)} Some agree well with the dispersive values while others
do so only for large values of $L_4^r,L_6^r$. There is also no obvious
pattern to which values for $L_4^r,L_6^r$ gives the best agreements.
Some of these difficulties could be due to the fact that the lowest order
is small in the region relevant for the subthreshold expansion as
can be seen in Figs.~\ref{fig:t120} and \ref{fig:t320}.

We now discuss some of them to show these issues also
considering
their agreement with the dispersive calculation. The latter can also
be judged from Tab.~\ref{tab:subthres2}.
The $c^+_{00}$ component agrees with the dispersive result in a very small
region for large negative $L_4^r$. That was precisely the place where
the $\chi^2$ of the fits for the input parameters started getting large.
$c^+_{20}$ agrees excellently at order $p^4$ but gets fairly large $p^6$
corrections. It starts agreeing once more 
for larger positive values of $L_4^r$ and $L_6^r$
than considered here. $c^-_{10}$ had a large negative estimate
of the contribution from the $p^6$ constants. This drives the total
$p^6$ contribution to be negative and the total result stays between 0.04
and 0.15, significantly below the dispersive result of \cite{BDM}.
The remaining subthreshold parameters all have large $p^6$ corrections and
it is not clear whether we have a convergent series or not. But the
general size and the sign is correct.

We now turn to the scattering lengths. The kinematical quantities
here have values which are already large for ChPT but looking at
Figs.~\ref{fig:t120} and \ref{fig:t320} the convergence seems fine in that
region. The finer features like higher partial waves and the radii
might however work less well.

\FIGURE{
\begin{minipage}{7cm}
\includegraphics[width=7cm,angle=270]{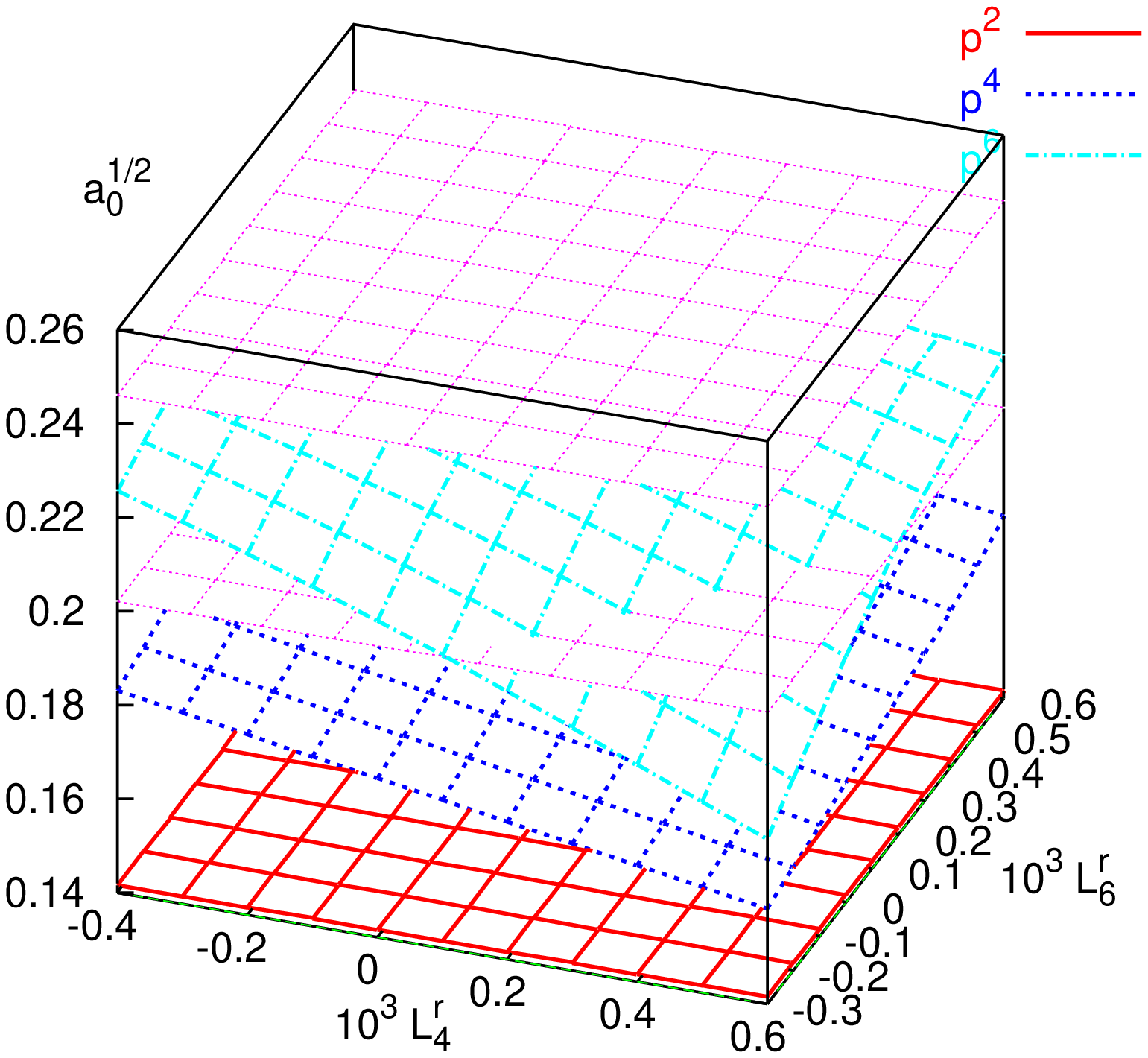}
\centerline{(a)}
\end{minipage}
\begin{minipage}{7cm}
\includegraphics[width=7cm,angle=270]{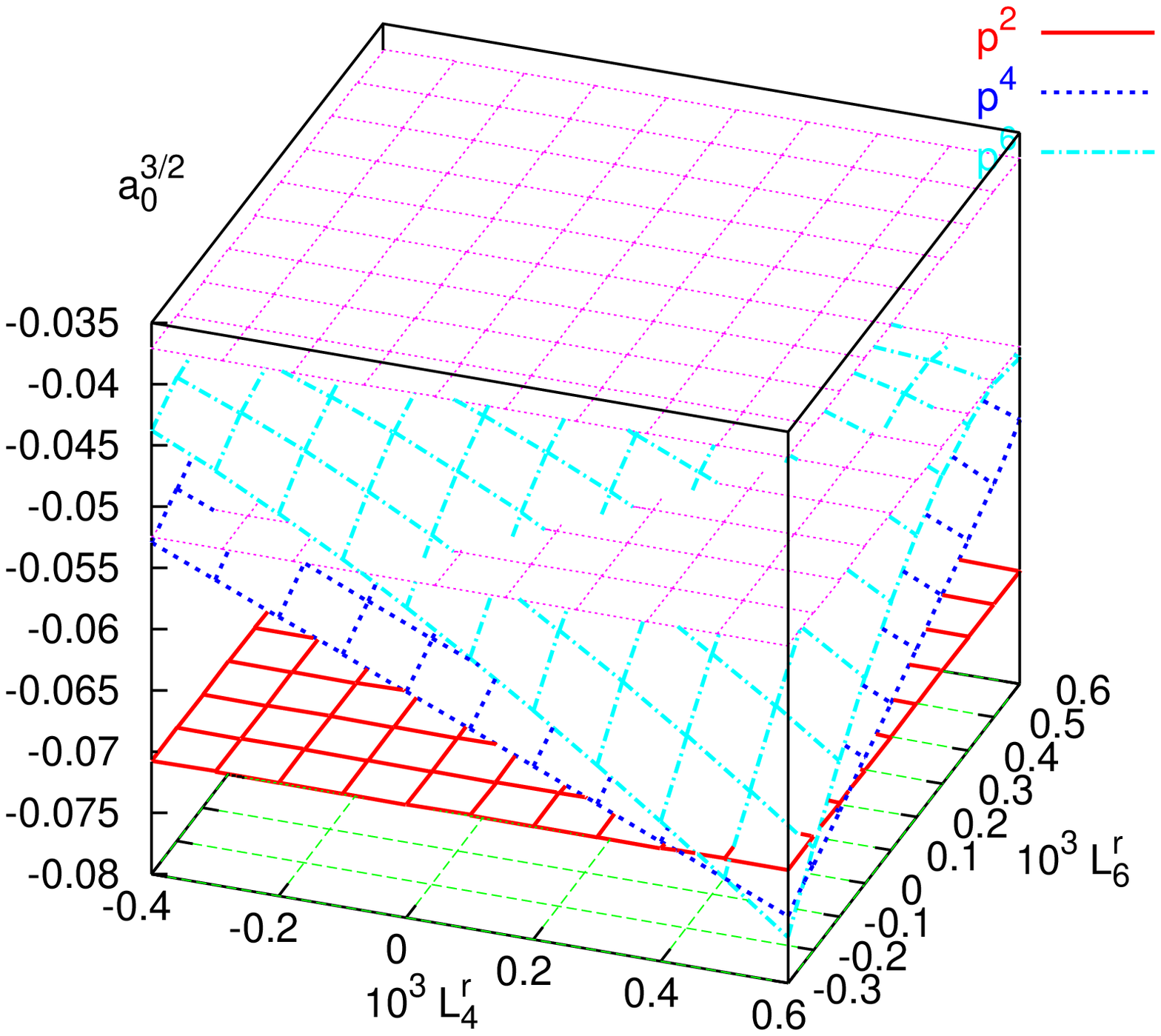}
\centerline{(b)}
\end{minipage}
\caption{\label{fig:a120}\label{fig:a320} 
(a) The scattering length $a^{1/2}_0$ as a function
of the input values of $L_4^r$ and $L_6^r$ used in the $L_i^r$ determination.
The unlabeled planes are the result of \cite{BDM} with their errors.
 (b) The same for the scattering length $a^{3/2}_0$.}
}
In Fig.~\ref{fig:a120}(a) we have plotted 
the $S$ wave scattering length in the isospin 1/2 channel. 
The series shows a nice
convergence and agrees with the dispersive result for most of the
$L_4^r$-$L_6^r$ region we considered. Only a small region of 
negative $L_6^r$ and positive $L_4^r$ disagrees.
The result for the $S$ wave scattering
length in the $I=3/2$ channel, shown in Fig,~\ref{fig:a320}(b),
has qualitatively the same behaviour, 
ruling out a somewhat larger region of the $(L_4^r,L_6^r)$ plane.
For the $P$ wave scattering lengths, we get agreement in the $I=1/2$ channel
with the dispersive
result in essentially the whole region considered with a preference for
positive values of $L_4^r$. The $I=3/2$ channel,
$a^{3/2}_1$ has large $p^6$ corrections always leading to a value
significantly above the dispersive result.
Looking at higher threshold parameters the picture is again mixed.  
$b^{1/2}_0$ is
typically 40 to 60\% above the dispersive result, $b^{3/2}_0$ is too
small by 20 to 50\%
and $b_1^{1/2}$ and $b^{3/2}_1$ have obvious convergence problems.
We have shown the results for fit 10 of \cite{ABT4} in Tab.~\ref{tab:aij2}
together with dispersive estimates of \cite{BDM}.

\TABLE{
\begin{tabular}{rrr}
\hline
  & Fit 10 & \cite{BDM}\\
\hline
$     a^{1/2}_0$ & 0.220 & $0.224\pm0.022$   \\
$10   a^{1/2}_1$ & 0.18  & $0.19\pm0.01$ \\  
$10   a^{3/2}_0$ & $-0.47$ & $-0.448\pm0.077$ \\
$10^2 a^{3/2}_1$ & 0.31 & $0.065\pm0.044$ \\  
$10   b^{1/2}_0$ & 1.3 & $0.85\pm0.04$ \\
$10   b^{3/2}_0$ & $-0.27$ & $-0.37\pm0.03$ \\
$T^+_{CD}$       & 2.11 & $3.90\pm1.50$ \\
\hline
\end{tabular}
\caption{\label{tab:aij2} The results
for the scattering lengths and ranges 
and the amplitude at the Cheng-Dashen point as well as the dispersive result.
The scattering lengths and ranges are given in units of $m_{\pi^+}$.}

}

The value of the amplitude at the Cheng-Dashen point is related to the kaon
sigma term. The dependence on $L_4^r$ and $L_6^r$ is shown in Fig.~\ref{figCD}
together with the dispersive result of \cite{BDM}. The corrections are
large over most of the region and can be compared with the direct
calculation of the sigma term shown in Fig.~9(b) of \cite{BD}.

The overall picture we thus obtain in the end is rather mixed. If we look
only at the quantities, $c^+_{10}$, $c^-_{00}$, $a_0^{1/2}$ and
$a_0^{3/2}$ we see a series that converges reasonably well and
reasonable agreement with the dispersive result from \cite{BDM} is found
for large regions of the values in the  $(L_4^r,L_6^r)$ plane. In
particular the value with $(L_4^r,L_6^r)\sim (0,0)$, corresponding to fit 10
of \cite{ABT3}, is located inside the allowed region. The other quantities
present more difficult to interpret results depending on how one
judges their convergence and the (dis)agreement with the dispersive results of
\cite{BDM}.

\FIGURE{
\includegraphics[width=7cm,angle=270]{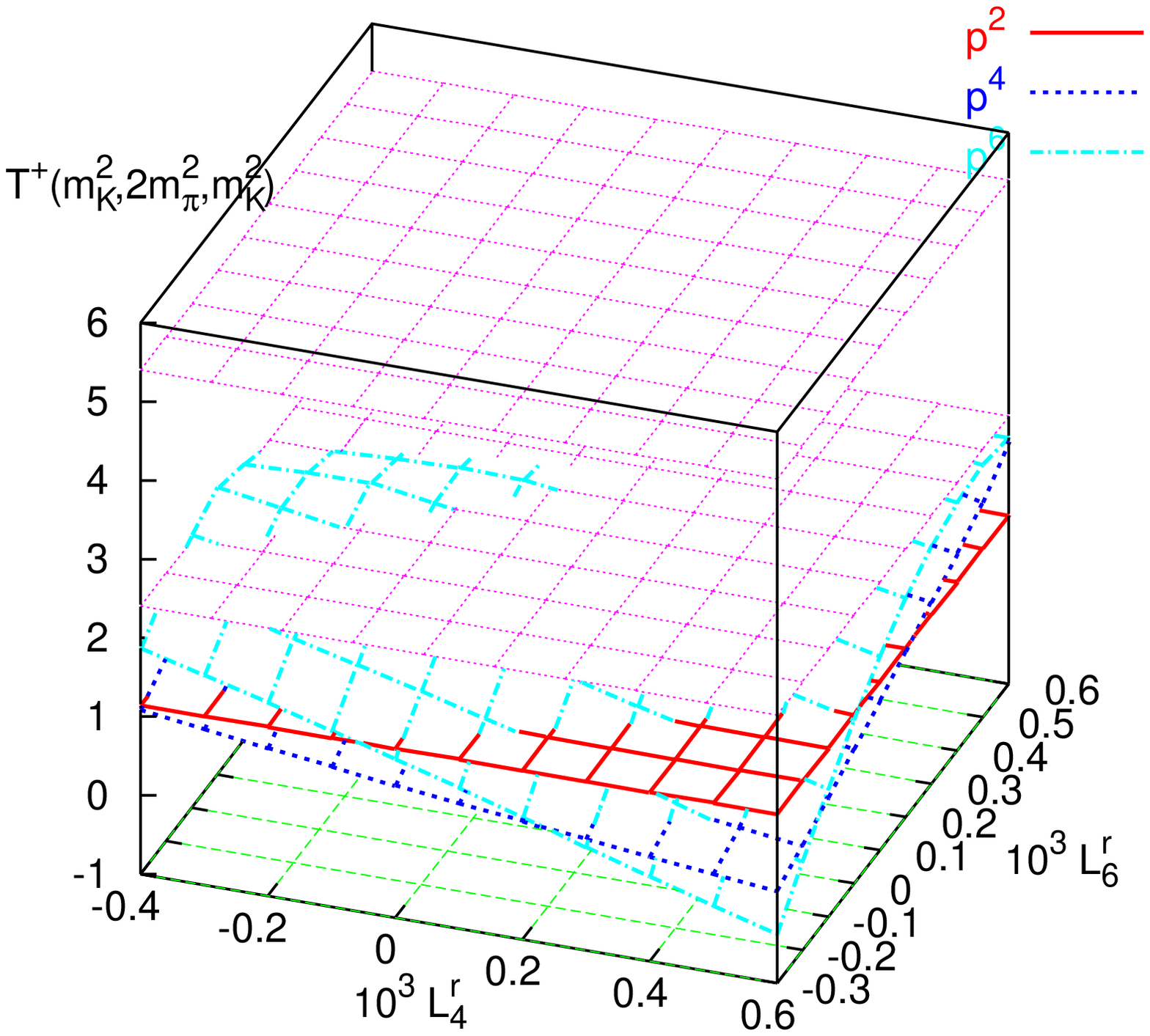}
\caption{\label{figCD} The amplitude at the Cheng-Dashen point as a function
of the input values of $L_4^r$ and $L_6^r$ together with the dispersive
estimates of \cite{BDM}.}
}

\subsection{$\pi K$, $\pi\pi$ and Scalar Form-factors}
\label{piKpipi}

Even though it seems that the $\pi K$ processes alone do not provide
strong restrictions on the low-energy constants
we can use it together with $\pi\pi$ scattering and the scalar form-factor
to restrict the region in the $(L_4^r,L_6^r)$ plane allowed.
A full analysis is planned for the future but here we discuss
the present results together with the earlier ones of \cite{BD,BDT}.

The constraints in the $(L_4^r,L_6^r)$ plane
come from several sources:

{\bf 1.}
The region inferred by the scalar form-factor analysis of \cite{BD}.
\begin{equation}
\label{ffl4l6}
L_6^r  \approx L_4^r - 0.35 \, 10^{-3}\,.
\end{equation}
This came from two arguments: 
(i) The assumption that the pion and kaon isoscalar
scalar form-factors at zero
do not deviate by large factors from their lowest order values, as
judged from Figs.~9(a) and 9(b) in \cite{BD}.
(ii) The agreement of the ChPT calculation of the pion scalar radius
with the dispersive results. Notice that the dispersive results used
the values of the form-factors at zero as input. The ChPT prediction
for the radius alone was rather constant as shown in Fig.~11(a) in
\cite{BD}.

{\bf 2.} From $\pi\pi$ scattering \cite{BDT} we got constraints from
four sources. The parameters $C_1$ and $C_2$ of \cite{CGL2} similar to
subthreshold parameters and the scattering lengths $a^0_0$ and $a^2_0$.
The constraints are
{\bf (a)} $a_0^{0}$ agrees reasonably well over the whole
region considered, see Fig.~4(a) in \cite{BDT}.
{\bf (b)} $a_0^{2}$ gives the strongest constraints,
see Fig.~4(b) in \cite{BDT}. It requires 
\be
L_4^r\gtrsim 0.27\,10^{-3}+0.9\,L_6^r\,.
\ee
{\bf (c)} $C_1$ imposes $ L_4^r \gtrsim 0.2~10^{-3}$ and does
not constrain $L_6^r$,
see Fig.~5(a) in \cite{BDT}.
{\bf (d)} $C_2$ does not provide significant more constraints than the above,
note that the plot shown in \cite{BDT} is erroneous.

{\bf 3.} We review now the constraints we found in this article
from  $\pi K$ scattering.
{\bf (a)} $c_{00}^-$ imposes no constraint, see Fig.~\ref{fig:cm00}(b).
{\bf (b)} $c_{10}^+$, delimits 
\be 
L_4^r \lesssim 0.28 \, 10^{-3}+0.25\,L_6^r\,,
\ee
see Fig.~\ref{fig:cp10}(a).
{\bf (c)} The constraint from $a_0^{1/2}$ is contained in the $a_0^{3/2}$,
see Fig.~\ref{fig:a120}(a). 
{\bf (d)} The constraint from $a_0^{3/2}$, see Fig.~\ref{fig:a320}(b),
is $L_6^r \gtrsim L_4^r - 0.4 \, 10^{-3}$.

Here we have used an error double the ones quoted in the
dispersive results of \cite{CGL2,BDM} to take
the convergence of the chiral series somewhat into account as well.

The $\pi K$ constraint is essentially only from $c^+_{10}$,
while for $\pi\pi$ scattering it is mainly $a^2_0$ with a little extra
from $C_1$. 
These constraints do not overlap but the various regions almost touch
for $L_4^r\approx 0.2\,10^{-3}$ and $L_6^r\lesssim 0.0\,10^{-3}$.
At present the situation is only {\em marginally compatible}.

With the uncertainties associated in the calculations:
the estimates of ${\cal O}(p^6)$ LECs,
correlations between the $L_i^r$ and
the fact that the errors on the $L_i^r$ from $K_{\ell4}$, the masses and
decay constants is not yet taken into account the above
conclusion is preliminary.

\section{Conclusions}
\label{conclusions}

In this paper we have calculated the $\pi K$ scattering amplitude to
next-to-next-to-leading order. We have presented analytically the results
to next-to-leading order and the dependence on the $p^6$ LECs $C_i^r$.
The remaining analytical expressions at order $p^6$ are very long and can
be obtained from \cite{formulas} or from the authors. This calculation is the
main result of this work.

We presented some numerics with the LECs at order $p^4$ and
$p^6$ set equal to zero at the scale of the rho mass. These results
allow a first impression about the convergence of the series for various
quantities, in particular we have presented results for the
subthreshold parameters, the scattering lengths and the amplitude at
the Cheng-Dashen point.

The second part of this work was a first attempt of extending the order
$p^4$ work of constraining low-energy constants from $\pi K$ scattering
of \cite{BDM,AB,ABM}. 
In these works values for $L_4^r$ were suggested
that are positive and different from zero. 
We used in this work as inputs the correlated values
for the $L_i^r$ determined from the $K_{\ell4}$ form-factors, pseudoscalar
meson masses and decay constants and  estimated the contributions from
the order $p^6$ constants 
with the same procedure as was used in previous next-to-next-to-leading
processes. 
In contrast to the $p^4$ results, a first estimate for
the quantities which appear to be most reliably
obtainable from the $\pi K$ scattering amplitude at order $p^6$ are fully
compatible with both the $1/N_c$ suppressed LECs $L_4^r$ and $L_6^r$ being
equal to zero. Contrary to expectations, the study of
these  $\pi K$ (sub)threshold parameters alone do not allow
to draw definite conclusions on the presence of large Zweig rule violating
contributions as discussed in \cite{Stern2,Stern3} and references therein.

The remaining $\pi K$ quantities present a rather mixed picture, the
convergence of the series is often questionable and the agreement with
the results from the dispersive analysis is at the same level but no
obvious large discrepancies exist. The estimated contribution of the $p^6$
constants to many quantities is fairly large and rather uncertain,
especially those involving the scalars.

We have also studied how these results fit together with the earlier
ones on $\pi\pi$ scattering and the scalar form-factors. We found only marginal
compatibility as described in Sect.~\ref{piKpipi}.

Planned work for the future is to combine
all existing order $p^6$ calculations in three flavour
ChPT in order to determine from experiment and/or dispersion theory
as many as possible of the $p^6$ constants and to fully take into account
all correlations for the $L_i^r$ and errors on the experimental and
dispersive inputs used.

\acknowledgments

The program FORM 3.0 has been used extensively in these calculations
\cite{FORM}. J.B. and P.D.
are supported in part by the Swedish Research Council
and European Union TMR
network, Contract No. HPRN-CT-2002-00311  (EURIDICE). P.T acknowledges
support by the Spanish Research Council.


\begin{thebibliography}{99}

\bibitem{Weinberg}
S.~Weinberg,
Physica A {\bf 96} (1979) 327.

\bibitem{GL1}
J.~Gasser and H.~Leutwyler,
Annals Phys.\  {\bf 158} (1984) 142.

\bibitem{GL2}
J.~Gasser and H.~Leutwyler,
Nucl.\ Phys.\ B {\bf 250} (1985) 465.

\bibitem{Bijnenskl4}
J.~Bijnens,
Nucl.\ Phys.\ B {\bf 337} (1990) 635.

\bibitem{Riggenbach}
C.~Riggenbach, J.~Gasser, J.~F.~Donoghue and B.~R.~Holstein,
Phys.\ Rev.\ D {\bf 43} (1991) 127.

\bibitem{Stern1}
J.~Stern, H.~Sazdjian and N.~H.~Fuchs,
Phys.\ Rev.\ D {\bf 47} (1993) 3814
[arXiv:hep-ph/9301244].

\bibitem{BCEGS1}
J.~Bijnens, G.~Colangelo, G.~Ecker, J.~Gasser and M.~E.~Sainio,
Phys.\ Lett.\ B {\bf 374} (1996) 210
[arXiv:hep-ph/9511397].

\bibitem{BCEGS2}
J.~Bijnens, G.~Colangelo, G.~Ecker, J.~Gasser and M.~E.~Sainio,
Nucl.\ Phys.\ B {\bf 508} (1997) 263
[Erratum-ibid.\ B {\bf 517} (1998) 639]
[arXiv:hep-ph/9707291].

\bibitem{BCT}
J.~Bijnens, G.~Colangelo and P.~Talavera,
JHEP {\bf 9805} (1998) 014
[arXiv:hep-ph/9805389].

\bibitem{ACGL}
B.~Ananthanarayan, G.~Colangelo, J.~Gasser and H.~Leutwyler,
Phys.\ Rept.\  {\bf 353} (2001) 207
[arXiv:hep-ph/0005297].

\bibitem{CGL1}
G.~Colangelo, J.~Gasser and H.~Leutwyler,
Phys.\ Lett.\ B {\bf 488} (2000) 261
[arXiv:hep-ph/0007112].

\bibitem{CGL2}
G.~Colangelo, J.~Gasser and H.~Leutwyler,
Nucl.\ Phys.\ B {\bf 603} (2001) 125
[arXiv:hep-ph/0103088].

\bibitem{Pislak1}
S.~Pislak {\it et al.}  [BNL-E865 Collaboration],
Phys.\ Rev.\ Lett.\  {\bf 87} (2001) 221801
[arXiv:hep-ex/0106071].

\bibitem{Pislak2}
S.~Pislak {\it et al.},
Phys.\ Rev.\ D {\bf 67} (2003) 072004
[arXiv:hep-ex/0301040].

\bibitem{Stern2}
S.~Descotes-Genon, L.~Girlanda and J.~Stern,
Eur.\ Phys.\ J.\ C {\bf 27} (2003) 115
[arXiv:hep-ph/0207337].

\bibitem{Stern3}
S.~Descotes-Genon, N.~H.~Fuchs, L.~Girlanda and J.~Stern,
arXiv:hep-ph/0311120.

\bibitem{LJP}
L.~Girlanda, J.~Stern and P.~Talavera,
Phys.\ Rev.\ Lett.\  {\bf 86} (2001) 5858
[arXiv:hep-ph/0103221].

\bibitem{BDT}
J.~Bijnens, P.~Dhonte and P.~Talavera,
JHEP {\bf 0401} (2004) 050
[arXiv:hep-ph/0401039].

\bibitem{BD}
J.~Bijnens and P.~Dhonte,
JHEP {\bf 0310} (2003) 061
[arXiv:hep-ph/0307044].

\bibitem{ABT1}
G.~Amoros, J.~Bijnens and P.~Talavera,
Nucl.\ Phys.\ B {\bf 568} (2000) 319
[arXiv:hep-ph/9907264].

\bibitem{Post1}
P.~Post and K.~Schilcher,
Nucl.\ Phys.\ B {\bf 599} (2001) 30
[arXiv:hep-ph/0007095].

\bibitem{Post2}
P.~Post and K.~Schilcher,
Eur.\ Phys.\ J.\ C {\bf 25} (2002) 427
[arXiv:hep-ph/0112352].

\bibitem{BT1}
J.~Bijnens and P.~Talavera,
JHEP {\bf 0203} (2002) 046
[arXiv:hep-ph/0203049].

\bibitem{BT2}
J.~Bijnens and P.~Talavera,
Nucl.\ Phys.\ B {\bf 669} (2003) 341
[arXiv:hep-ph/0303103].

\bibitem{Weinbergpipi}
S.~Weinberg,
Phys.\ Rev.\ Lett.\  {\bf 17} (1966) 616.

\bibitem{Griffith}
R.W. Griffith,
Phys. Rev. 176 (1968) 1705.

\bibitem{Langreview}
C.~B.~Lang,
Fortsch.\ Phys.\  {\bf 26} (1978) 509.

\bibitem{Lang}
C.~B.~Lang and W.~Porod,
Phys.\ Rev.\ D {\bf 21} (1980) 1295.

\bibitem{JN} 
N.~Johannesson and G.~Nilsson,
Nuovo Cim.\ A {\bf 43} (1978) 376.

\bibitem{BKM}
V.~Bernard, N.~Kaiser and U.~G.~Meissner,
Nucl.\ Phys.\ B {\bf 357} (1991) 129.

\bibitem{BKM2}
V.~Bernard, N.~Kaiser and U.~G.~Meissner,
Phys.\ Rev.\ D {\bf 43} (1991) 2757.

\bibitem{SaBorges1}
J.~Sa Borges, J.~Soares Barbosa and V.~Oguri,
Phys.\ Lett.\ B {\bf 412} (1997) 389.

\bibitem{SaBorges2}
J.~Sa Borges and F.~R.~A.~Simao,
Phys.\ Rev.\ D {\bf 53} (1996) 4806.

\bibitem{BKM3}
V.~Bernard, N.~Kaiser and U.~G.~Meissner,
Nucl.\ Phys.\ B {\bf 364} (1991) 283.

\bibitem{JOP}
M.~Jamin, J.~A.~Oller and A.~Pich,
Nucl.\ Phys.\ B {\bf 587} (2000) 331
[arXiv:hep-ph/0006045].

\bibitem{MO}
U.~G.~Meissner and J.~A.~Oller,
Nucl.\ Phys.\ A {\bf 679} (2001) 671
[arXiv:hep-ph/0005253].

\bibitem{Roessl}
A.~Roessl,
Nucl.\ Phys.\ B {\bf 555} (1999) 507
[arXiv:hep-ph/9904230].

\bibitem{FKM}
M.~Frink, B.~Kubis and U.~G.~Meissner,
Eur.\ Phys.\ J.\ C {\bf 25} (2002) 259
[arXiv:hep-ph/0203193].

\bibitem{AB}
B.~Ananthanarayan and P.~Buttiker,
Eur.\ Phys.\ J.\ C {\bf 19} (2001) 517
[arXiv:hep-ph/0012023].

\bibitem{BDM}
P.~Buttiker, S.~Descotes-Genon and B.~Moussallam,
arXiv:hep-ph/0310283.

\bibitem{Nehme1}
A.~Nehme,
Eur.\ Phys.\ J.\ C {\bf 23} (2002) 707
[arXiv:hep-ph/0111212].

\bibitem{Nehme2}
A.~Nehme and P.~Talavera,
Phys.\ Rev.\ D {\bf 65} (2002) 054023
[arXiv:hep-ph/0107299].

\bibitem{KM1}
B.~Kubis and U.~G.~Meissner,
Phys.\ Lett.\ B {\bf 529} (2002) 69
[arXiv:hep-ph/0112154].

\bibitem{KM2}
B.~Kubis and U.~G.~Meissner,
Nucl.\ Phys.\ A {\bf 699} (2002) 709
[arXiv:hep-ph/0107199].

\bibitem{ABT4}
G.~Amoros, J.~Bijnens and P.~Talavera,
Nucl.\ Phys.\ B {\bf 602} (2001) 87
[arXiv:hep-ph/0101127].

\bibitem{chptlectures}
A.~Pich, Lectures at Les Houches Summer School in
Theoretical Physics, Session 68: Probing the Standard Model of Particle
Interactions, Les Houches, France, 28 Jul - 5 Sep 1997,
[hep-ph/9806303];\\
G.~Ecker,
Lectures given at Advanced School on Quantum Chromodynamics (QCD 2000),
Benasque, Huesca, Spain, 3-6 Jul 2000,
[hep-ph/0011026];\\
S.~Scherer,
hep-ph/0210398.

\bibitem{BCE}
J.~Bijnens, G.~Colangelo and G.~Ecker,
JHEP {\bf 9902} (1999) 020
[arXiv:hep-ph/9902437].

\bibitem{BCE2}
J.~Bijnens, G.~Colangelo and G.~Ecker,
Annals Phys.\  {\bf 280} (2000) 100
[arXiv:hep-ph/9907333].

\bibitem{BCE3}
J.~Bijnens, G.~Colangelo and G.~Ecker,
Phys.\ Lett.\ B {\bf 441} (1998) 437
[arXiv:hep-ph/9808421].

\bibitem{formulas}
These can be downloaded
from \verb;http://www.thep.lu.se/~bijnens/chpt.html;.

\bibitem{Ecker1}
G.~Ecker, J.~Gasser, A.~Pich and E.~de Rafael,
Nucl.\ Phys.\ B {\bf 321} (1989) 311.

\bibitem{Ecker2}
G.~Ecker, J.~Gasser, H.~Leutwyler, A.~Pich and E.~de Rafael,
Phys.\ Lett.\ B {\bf 223} (1989) 425.

\bibitem{ABT3}
G.~Amoros, J.~Bijnens and P.~Talavera,
Nucl.\ Phys.\ B {\bf 585} (2000) 293
[Erratum-ibid.\ B {\bf 598} (2001) 665]
[arXiv:hep-ph/0003258].

\bibitem{Pich}
V.~Cirigliano, G.~Ecker, H.~Neufeld and A.~Pich,
JHEP {\bf 0306} (2003) 012
[arXiv:hep-ph/0305311].

\bibitem{BCG}
J.~Bijnens, G.~Colangelo and J.~Gasser,
Nucl.\ Phys.\ B {\bf 427} (1994) 427
[arXiv:hep-ph/9403390].

\bibitem{BGLP}
J.~Bijnens, E.~Gamiz, E.~Lipartia and J.~Prades,
JHEP {\bf 0304} (2003) 055
[arXiv:hep-ph/0304222].

\bibitem{MHA}
M.~Knecht and A.~Nyffeler,
Eur.\ Phys.\ J.\ C {\bf 21} (2001) 659
[arXiv:hep-ph/0106034];\\
S.~Peris, M.~Perrottet and E.~de Rafael,
JHEP {\bf 9805} (1998) 011
[arXiv:hep-ph/9805442].

\bibitem{ABM}
B.~Ananthanarayan, P.~Buettiker and B.~Moussallam,
Eur.\ Phys.\ J.\ C {\bf 22} (2001) 133
[arXiv:hep-ph/0106230].

\bibitem{FORM}
J.~A.~Vermaseren,
math-ph/0010025.



\end{thebibliography}
\end{document}